\documentclass[aps,twocolumn,groupedaddress,prl,letter]{revtex4}
\usepackage{epsfig}

\begin{document}

\title{Demonstration of photomultiplier tube operation at 29~K}

\author{James A. Nikkel}
\email{james.nikkel@yale.edu}
\author{W. Hugh Lippincott}
\author{Daniel N. McKinsey}
\affiliation{Yale University, New Haven, CT}

\date{\today}

\begin{abstract}

We describe measurements of gain, dark current, and quantum efficiency
obtained while cooling a Hamamatsu R5912-02-MOD 
photomultiplier tube from room temperature to 29~K.  We found that the PMT operated
normally down to 29~K, with a reduced gain and quantum efficiency at the
lowest temperatures.  Furthermore, we found that the dark count rate increased
as the temperature decreased.  We conclude that these PMTs appear to be
adequate for the requirements of the CLEAN experiment.  

\end{abstract}

\maketitle

\section{Introduction}

The CLEAN detector is designed to collect scintillation light from solar neutrinos
and dark matter particles interacting in liquid
neon~\cite{McKinsey:2005b}. The engineering design of CLEAN is greatly
simplified by operating the photomultiplier tubes (PMTs) directly in
liquid neon at 27~K.  While there are experiments that use PMTs in liquid  
xenon (165~K)~\cite{Mihara:2004, Aprile:2005a}, liquid argon
(87~K)~\cite{Arneodo:2001}, and liquid nitrogen (77~K)~\cite{Ankowski:2006},
there are no published measurements of the characteristics of PMTs below 77~K.   

In this paper we present measurements of the gain, dark
count rate, and quantum efficiency of a Hamamatsu R5912-02-MOD PMT as a
function of temperature between 300~K and 29~K. This PMT has fourteen dynodes,
four more than the standard R5912, and it has a platinum underlay on the 
photocathode to improve performance at low temperatures. 
When standard PMTs are operated at low temperatures, the electrical
conductivity of the photocathode decreases as the photocathode is cooled,
resulting in space charge buildup and a reduced quantum efficiency.  The
platinum underlay corrects this situation by allowing charge to flow to any
point on the photocathode regardless of the photocathode conductivity.  The
disadvantage is a small loss of light transmission due to the extra layer
of material on the PMT face.  According to manufacturer specifications,
the room temperature quantum efficiency of the Hamamatsu R5912
decreases from 22\% to about 16\% at 390~nm after the addition of the platinum
layer~\footnote{Norm Schiller, Hamamatsu, personal communications.}.

\section{Experimental Details}
\label{sec:exp_det}

\begin{figure}[h]
  \centerline{
    \hbox{\psfig{figure=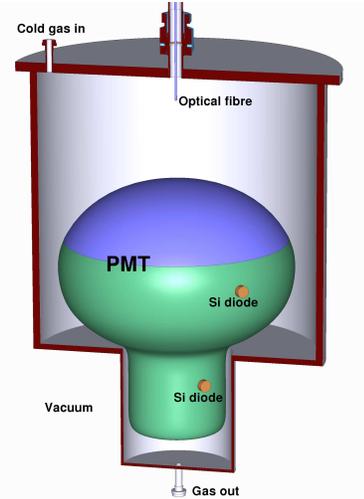,width=5cm,%
                clip=}}}
          \caption[Scintillation cell]
                  {Schematic representation of the PMT test cell.}
          \label{fig:cell}
\end{figure}

Figure~\ref{fig:cell} is a schematic of the test cell used to cool and
illuminate the PMT.  The vessel was a stainless steel can with 
feedthroughs for the PMT signal, PMT high voltage, thermometers, gas service,
and an optical fibre.  The incoming gas was cooled by a pulse tube
refrigerator~\footnote{Cryomech, model PT805.}, and the temperature of the PMT was monitored by
direct readout of two silicon diodes~\footnote{Lakeshore, model DT470-CU-11a.} attached to its exterior with thermal
grease.  We used dry nitrogen exchange gas at 1100~mbar for temperatures between room temperature and
100~K, and neon gas at 1500~mbar for temperatures below 100~K. We used nitrogen at
high temperatures because neon can potentially diffuse through
the PMT glass causing after-pulsing in the tube, as has been observed in  
helium~\cite{Bartlett:1981}.  By introducing the neon gas only below 100~K, the
diffusion rate was sufficiently suppressed to eliminate this risk. 

The cooling and heating rates were kept below 2.5~K/hour between 100~K and
300~K, and below 4.5~K/hour between 29~K and 100~K.  The vessel was
kept at 29~K for approximately 9 days before warming.  

We introduced from the top of the cell a single optical fibre, ground to a
point to illuminate as much of the photocathode surface as possible. 
One end pointed towards the PMT while the other end of the fibre was
illuminated by an LED pulser at room temperature.  This pulser consisted of four
blue LEDs~\footnote{Fairchild Semiconductor, model MV8B01.} 
with a peak wavelength of 430~nm glued into an acrylic block mounted
to a photodiode.  The fibre connectors were mounted to the side of the block so that
only a small fraction of the light illuminating  the photodiode
entered the fibre and reached the PMT.  The photodiode monitored the gross
light output from the LEDs to ensure consistent illumination of the PMT.

We used a voltage divider design recommended by Hamamatsu for the R5912-02-MOD
PMT with separate cables for high voltage and signal.  The capacitors and
resistors for the divider were individually tested in liquid nitrogen to check for
temperature-dependent variation before assembly.  The PMT anode voltage was set to
+1500~V.    

The LEDs were powered by a digital pulse generator~\footnote{Berkeley Nucleonics, model
555.} and their light output was monitored with the photodiode.  The LEDs were
pulsed with either 5.8~V or 6.2~V at a rate of 20~Hz and a pulse width of $32~\mu$s. 
The coupling between the LED pulser and the fibre was weak enough that the
signal from the PMT consisted of discrete single photoelectron events spread
over the duration of the LED pulse.

\section{Experimental Results}



\begin{figure}[h]
  \centerline{
    \hbox{\psfig{figure=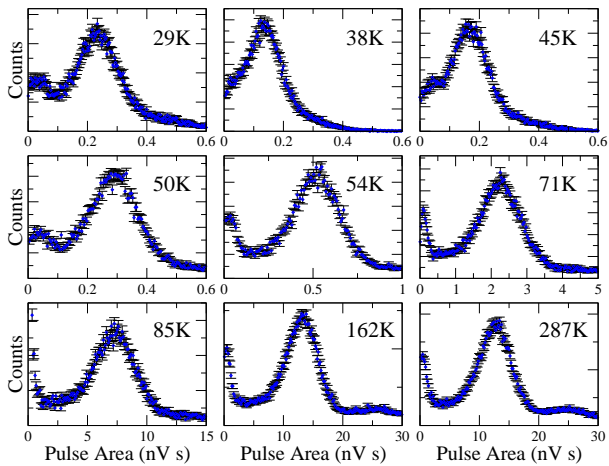,width=8cm,%
                clip=}}}
          \caption[Singles peaks]
                  {Plots of the single photoelectron spectra at various
                    temperatures.}
          \label{fig:singles}
\end{figure}

\begin{figure}[h]
  \centerline{
    \hbox{\psfig{figure=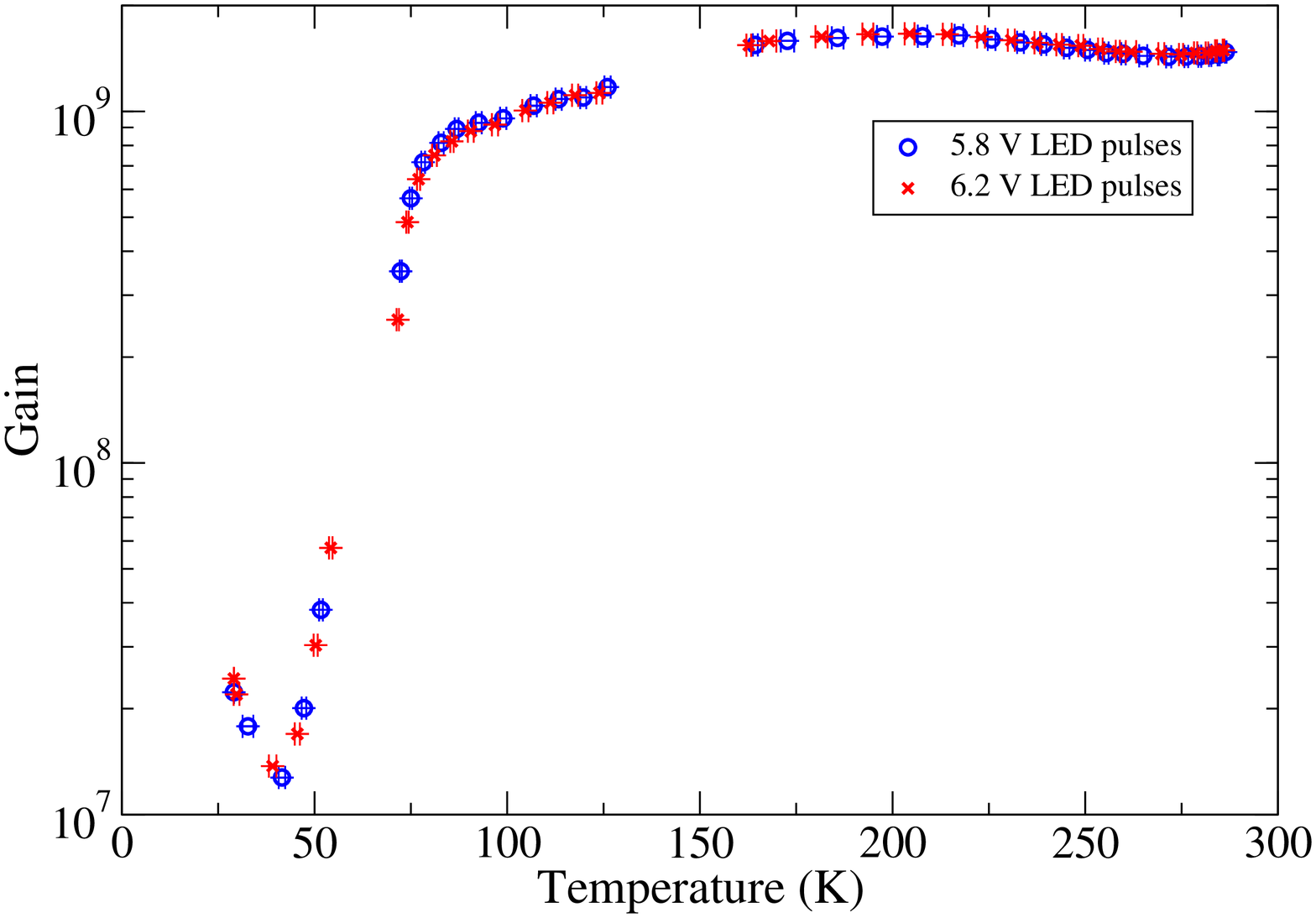,width=8cm,%
                clip=}}}
          \caption[Gain curves]
                  {Plot of the single photoelectron peak position
                    vs. temperature.}
          \label{fig:gain}
\end{figure}

By measuring the value of the single photoelectron peaks, we were able to
obtain  the PMT gain as a function of temperature.  The PMT voltage trace was
divided into 150~ns regions where the trace crossed an experimentally
determined threshold of about 1/5 of a photoelectron. These regions were
integrated to obtain the pulse area.  Fig.~\ref{fig:singles} shows a series of
histograms of the pulse area at various temperatures.  By making Gaussian fits
to the peaks, we obtained curves for the absolute gain
as a function of temperature, shown in Fig.~\ref{fig:gain}. 

There is currently no quantitative model that explains all of the structure
seen in Fig.~\ref{fig:gain}.  Possible explanations include
thermal contraction of the PMT components or a variation in emissivity of the
dynodes.  Measured changes in the voltage divider component values were much too
small to account for the observed variation of the gain.


To determine the dark count rate, we counted the number of single photoelectrons in a fixed
interval while the PMT was not externally illuminated.  The single photoelectron spectra were
integrated and rounded off into integer multiples of the single photoelectron
peak area for counting, then divided by the LED pulse length to obtain the
rate. At room temperature, we measured a dark count rate of approximately 300 counts per second
(cps). 

\begin{figure}[h]
  \centerline{
    \hbox{\psfig{figure=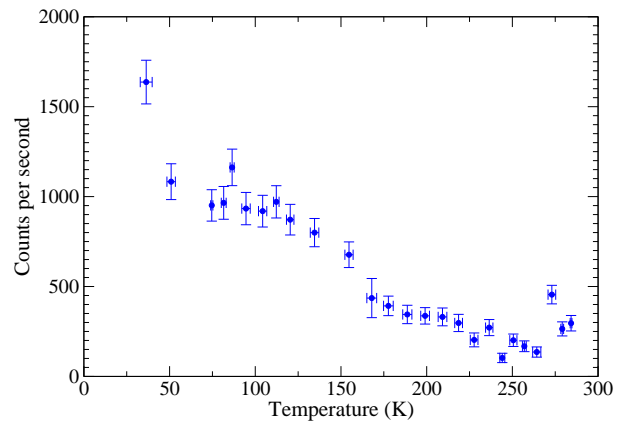,width=8cm,%
                clip=}}}
          \caption[Dark count rate]
                  {Plot of the dark count rate as a function of temperature.}
          \label{fig:QE}
\end{figure}

Fig.~\ref{fig:QE} is a plot of  the dark count rate as a function
of temperature. The uncertainties are dominated by photoelectron counting
statistics.  We think that the contribution from scintillation in the gases was
low as the nitrogen and neon were swapped at 100~K with less than a 14\%
change in the dark count rate. 


\begin{figure}[h]
  \centerline{
    \hbox{\psfig{figure=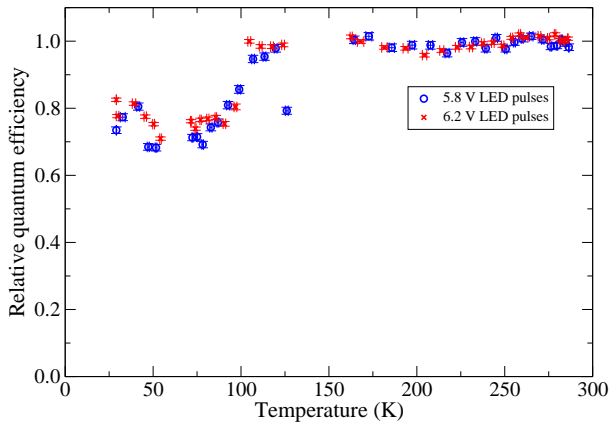,width=8cm,%
                clip=}}}
          \caption[QE]
                  {Plot of the quantum efficiency as a function of temperature relative to
                    the room temperature value.}
          \label{fig:QE}
\end{figure}

The PMT quantum efficiency was estimated by counting the number of
photoelectrons observed during the $32~\mu$s LED pulse duration. At room
temperature, the count rates were $6.5\times 10^4$~cps and 
$2.8\times 10^5$~cps when the LEDs were powered at 5.8 V and 6.2 V
respectively.  The repetition rate was kept at 20~Hz so the room temperature
time average for the pulsed signal was 42~cps and 170~cps above the dark rate
for 5.8~V and 6.2~V respectively. 

To check for depletion in the second half of the pulse, we
compared the first and last $16~\mu$s of each pulse.
The ratio of signal in the first half of each pulse to the total signal in the
pulse was measured to be $0.504 \pm 0.009$ and $0.48 \pm 0.01$ at room
temperature and 29~K respectively, suggesting the absence of charge depletion over the
duration of individual pulses at both temperature extremes.  

As we did not know the absolute photon flux onto the PMT,
we measured the quantum efficiency relative to the room temperature value,
shown in  Fig.~\ref{fig:QE}.   We do not have a definitive explanation 
for the change in the measured quantum efficiency at 100~K.

This work was supported by the David and Lucile Packard Foundation.

\end{document}